\def\year{2015}
\documentclass[letterpaper]{article}
\usepackage{aaai}
\usepackage{times}
\usepackage{helvet}
\usepackage{courier}

\usepackage{balance}  
\usepackage{graphicx} 
\usepackage{multirow}
\usepackage{subfigure}
\usepackage{amsmath}

\newcommand{\omt}[1]{}
\newcommand{\xhdr}[1]{\paragraph*{\bf #1.}}

\begin{document}
%
\title{Coordination and Efficiency in Decentralized Collaboration}
\author{Daniel M. Romero \\ University of Michigan \\drom@umich.edu \And Dan Huttenlocher \\ Cornell University \\huttenlocher@cornell.edu  \And
Jon Kleinberg \\ Cornell University \\kleinber@cs.cornell.edu }
\maketitle

\begin{abstract}
Environments for decentralized on-line collaboration are now widespread 
on the Web, underpinning open-source efforts,
knowledge creation sites including Wikipedia, and other
experiments in joint production.
When a distributed group works together in such a setting,
the mechanisms they use for {\em coordination} can play an 
important role in the effectiveness of the group's performance.

Here we consider the trade-offs inherent in coordination in these
on-line settings, balancing the benefits to collaboration with
the cost in effort that could be spent in other ways.
We consider two diverse domains that each contain a wide range
of collaborations taking place simultaneously --- Wikipedia and GitHub ---
allowing us to study how coordination varies across different projects.
We analyze trade-offs in coordination along two main dimensions,
finding similar effects in both our domains of study:
first we show that, in aggregate, high-status projects on these sites 
manage the coordination trade-off at a different level than typical projects;
and second, we show that projects use a different balance of coordination
when they are ``crowded,'' with relatively small size but many participants.
We also develop a stylized theoretical model for the cost-benefit trade-off 
inherent in coordination and show that it qualitatively matches the
trade-offs we observe between crowdedness and coordination.
\end{abstract}

\section{Introduction}
In many settings on the Web, groups of people who may have
no off-line associations with one another 
come together around a project-oriented 
site that supports remote interaction,
discussion, and the production of a shared work product.
This style of highly decentralized collaboration --- the participants
are geographically dispersed and may not interact with each other
outside the context of the site ---
is the driving force behind a range of large open-source projects
hosted on sites such as GitHub, as well as
knowledge creations sites including Wikipedia and recent experiments
in massively collaborative problem-solving.

One of the crucial questions that emerges,
as these forms of interaction become increasingly influential,
is to understand what makes them effective at a structural level,
and to characterize the properties associated with better outcomes.
To address this question, we draw on a conceptual framework that
has proven powerful in off-line domains --- analyzing
the effectiveness of teams through their
{\em coordination mechanisms}.  These mechanisms broadly include
the set of practices that
help a team organize its collaboration on a task, for 
dividing up shared work, setting
intermediate goals, and resolving disagreements
\cite{Barnard_38,Cooper_79,Foushee_82,Helmreich_86,Malone-Crowston90,Samer_2000}.

\xhdr{The Present Work: Trade-Offs in Coordination}

A fundamental property of coordination is the cost-benefit 
trade-off that it entails.
Coordination is beneficial, but it comes at a cost ---
the work that team members put into coordination could be used on the substance of the project itself
\cite{Becker,Entin_99,Kittur_2009,Macmillan_04}. 
Referring to the communication overhead inherent in coordination, 
Macmillan et al.~write, 
\begin{quote}
{\em ``Because communication is essential to team performance, effective team cognition has a communication overhead associated with the exchange of information among team members. Communication requires both time and cognitive resources, and, to the extent that communication can be made less necessary or more efficient, team performance can benefit as a result'' \cite{Macmillan_04}.}
\end{quote}

Our understanding of this cost-benefit trade-off comes largely from
the study of relatively small face-to-face teams, as in the research noted
above. But the trade-offs involved in coordination are equally or more
pronounced in on-line domains due to the limited ability of on-line teams to rely on less costly implicit coordination mechanisms 
\cite{Entin_99,Kittur_2009,Stout_96,Wang_91}, 
which often require shared mental models that 
are difficult to maintain on-line 
\cite{Salas_90,Wittenbaum_96}. 

These trade-offs have clear implications for
the design of these systems.
There is thus an opportunity to combine the work in the organizational
literature on coordination and its consequences for performance with the 
long line of work on coordination and its uses in on-line domains
\cite{Kittur_Kraut_2008,Kittur_2009,Kittur_2007,Krieger_2009,Malone-Crowston90}
and to evaluate the findings in the context of some of the most
active on-line collaborative settings.

\xhdr{Research Questions for Coordination}
An important question in the literature on off-line domains has been
to understand the possible levels of coordination that balance the trade-off
between cost overhead and performance gains, 
leading to the best team and individual performance. 
When we move to the context of large on-line projects, 
how do such projects manage this trade-off, and can we identify 
principles in how this balance is managed?

We approach these questions both through the development of an analytical
framework and through the study of large on-line datasets.
In particular, we focus on Wikipedia articles and
GitHub projects as two rich collaborative domains that share some
essential abstract properties:
\begin{itemize}
\item They each contain projects in which there is a primary work product
and also a channel for coordination among members of the project team.
\item Participation in projects has an interlocking structure, 
in the sense that the participants in one project may also be involved 
in others.
\item Certain projects may have a higher level of status or visibility
than others; on Wikipedia certain articles are featured, and on GitHub
certain projects can have a non-trivial number of watchers.
\end{itemize}
These ingredients also relate in interesting ways to the coordination
framework proposed by Malone and Crowston \cite{Malone-Crowston90},
who identify goals, activities, actors, and interdependencies
as the general components of coordination.

We organize our work around two central trade-offs in 
coordination:
\begin{itemize}
\item Do high-status projects manage the coordination trade-off differently
from typical projects?
\item How does coordination relate to a project's team composition 
and crowdedness --- in particular, 
the amount of work produced relative to the number of members in the team?
\end{itemize}

In addition to the data analysis supporting these two questions, we
develop a mathematical model to capture the balance of costs and benefits in 
coordination abstractly.

It is important to note a few features of our approach.
First, we organize our work around the aggregate analysis of
large datasets, and our findings are correspondingly 
oriented toward trade-offs at this cumulative level.
Our investigation is thus complementary to more fine-grained 
studies of individual projects and the specifics of their
coordination strategies.
Second, while our analysis is performed on two particular domains ---
Wikipedia and GitHub --- we seek to articulate a framework that can
be applied to on-line collaboration across many contexts.
With this in mind, we develop the core components of our approach ---
projects and shared work, coordination mechanisms, interaction 
across projects, and measures of status and visibility --- at
a general level, and illustrate how they can be applied across these
two different contexts to yield closely related findings.
We hope through
this alignment of common structure to suggest a set of principles that can be 
used more broadly.

Finally, we believe that our analysis of coordination raises a number
of possible suggestions for design, as we discuss further in what follows.
Applications supporting collaboration increasingly seek to steer groups
of users toward effective interaction, and coordination mechanisms can
be a powerful component of this process.
But the trade-offs inherent in coordination make clear that it can be
a non-trivial problem to determine whether a system should be guiding
a group toward more or less coordination in a given situation.
By understanding how levels of coordination naturally vary with the
visibility of a project and the 
crowdedness of the setting, we can establish principles by which 
coordination mechanisms can be tuned based on the underlying context.

\xhdr{Summary of Results}
We establish results on the trade-offs in coordination for
our main questions described above.
We first show that there are significant aggregate differences in
the way that coordination is used in the higher-status projects on
both Wikipedia and GitHub, relative to the use of coordination in 
typical projects.
This suggests that properties related to coordination can
be relevant to questions of performance and visibility in
our on-line setting.
For Wikipedia, where the set of participants in a single project is
often larger and more diverse, we delve further into the question by
looking at how much the effort and coordination is concentrated
on a small set of the most active participants ---
this connects to an argument due to Kittur and Kraut, who posit
that a balance of effort in on-line
tasks in which a few participants do most of the work can correspond
to a form of implicit coordination \cite{Kittur_Kraut_2008}.

We then explore the relationship between coordination 
and the composition of the project. We find that additional coordination 
is most useful when there are many team members and
the task is small, resulting in a crowded environment; it is most wasteful 
when the number of team members is small and the size of the task is large. 
As with our results on high-status projects, our findings here too
are consistent in Wikipedia and GitHub. 

We supplement this analysis with a
formal model for studying the trade-off between
the cost of explicit coordination and the benefits it brings to a
project. The predictions of the model also suggest that crowdedness
is a key parameter in the coordination trade-off.

\section{Wikipedia Data}

In this section and the next, we describe our two datasets ---
Wikipedia and GitHub.
We begin with Wikipedia, which is the larger and more complex of the two,
and the one where we are in some cases
able to compute correspondingly more complex functions of the data.

For our purposes, each Wikipedia article constitutes a project that is
produced by the set of users who edit it.
Wikipedia captures many of
the basic features that one sees in on-line collaboration
more generally, and for our purposes it also exhibits 
three desirable properties.
First, since each article constitutes a project with its own internal
life-cycle, we can observe the history of many projects in a common environment, 
exploring sources of variation among them.
Second, Wikipedia contains explicit markers of success and failure
for projects, including recognition of certain highly successful
articles. Finally, Wikipedia has well-developed mechanisms for explicit 
coordination, along with extensive records of coordination
for each article \cite{Viegas_2007,Keegan2012}. Our data contains approximately 3.4 million articles, each with a discussion page. This corresponds to the entire edit history of English Wikipedia up to April of 2007, developed as a resource by Crandall et al \cite{Crandall}.

We now describe how the basic ingredients of our framework manifest
themselves on Wikipedia: we first consider coordination mechanisms,
and then measures for the status and visibility of articles.

\xhdr{Coordination Mechanisms on Wikipedia}
We look at two kinds of interactions to measure coordination in Wikipedia: 
(i) discussion edits, and (ii) comments left on article edits. 
Each article on Wikipedia has a \emph{discussion page}
that is used to discuss issues related to the editing
of the article such as planning, resolving arguments, and enforcing
conventions \cite{Viegas_2007}. We use the number
of edits to discussion pages as a measure of how much effort editors 
spend explicitly coordinating. 
There is significant variation across articles
in the amount of discussion-page editing and in aggregate we will see that the variation points to overall differences
across different types of articles.

There is a second widely-used form of coordination:
When a user edits a Wikipedia article, she has the option of including
a comment where she can briefly explain the nature of the edit.
Leaving comments is often helpful for other editors
because the comments allow them to easily identify the kinds of
edits other users have contributed. Comments are similar to discussion-page edits in that
editors use them to communicate about the editing of the
article, but in contrast with discussion edits, comments are much
terser and thus tend to explain the nature of an edit 
without long discussions. In this sense,
we can think of comments as lying somewhere on the spectrum between
explicit and implicit coordination, and more implicit than 
discussion-page edits.

In some of the analysis we will consider both of these coordination
mechanisms, but at other points we will focus on discussion edits,
since they are the mechanism that allows for explicit coordination, including the ability to engage in back-and-forth 
interaction between multiple participants.

\xhdr{Use of Discussion Pages and Comments in Wikipedia}

While edit comments and discussion pages are meant to facilitate the
collaboration among Wikipedia editors to write high-quality articles,
people can use these tools for any purpose. For example, as is often
the case on the Web, it is possible that discussion
pages and comment could be used for spam or
other unintended purposes. If this were the case, measuring
coordination by the number of discussion edits and comments would be
misleading. To address this issue and to further understand how people
use discussion pages and comments, we read a small random sample of comments
and discussion sections from our data and 
manually categorize them according to their purpose.

Using a set of 100 randomly selected comments, we first construct a set of
categories of purposes. Then, we sample a new set of 100 randomly
selected comments and organize them into the determined
categories. We use the same procedure to categorize a randomly selected section from 100 discussion pages. Tables \ref{Comment_cat}
and \ref{Discussion_cat} show the categories and the number of
examples in each category for comments and discussion pages,
respectively. We observe that the majority of the sampled
comments and discussions are directly related to the writing of the
article. Any comment or discussion that was not relevant to the
editing of the article was placed in the "Other" category. Out of the
100 categorized comments and discussions, only 11 and 6 fell in the
"Other" category.

This analysis suggests that edit comments and discussion pages are indeed
mostly used by editors to coordinate their collaboration on Wikipedia.
Having established that these tools seem to be employed to facilitate
coordination, we use the number of comments and discussion edits as a
proxy for how much explicit coordination the editors perform.

\begin{table}[h!]
\centering
\begin{tabular}{ |c | c| }
  \hline
  \bf{Category}&  \bf{Num}\\
  \hline
  \hline
  Mentions section & 52\\
    \hline
  Reverted edit & 14\\
    \hline
  Minor edit & 19 \\
    \hline
  Added content & 14 \\
    \hline
    Removed content & 7 \\
    \hline
  Correction & 2 \\
    \hline
  Mentions other users & 14 \\
    \hline
  Other & 11 \\
    \hline
\end{tabular}
\caption{Categories of Wikipedia edit comments and number of examples in each category. Categories are based on a random sample of 100 comments. Comments can belong to more than one category.}
\label{Comment_cat}
\end{table}

\begin{table}[h!]
\centering
\begin{tabular}{|c|c|c|}
    \hline
    \multicolumn{2}{|c|}{\bf{Category}} & \bf{Num}\\
       \hline
              \hline
    \multirow{2}{*}{Justify}&Text edit & 14\\
     \cline{2-3}
    &Change metadata & 4\\
       \hline
    \multirow{4}{*}{Suggest Action}&Specific text edit & 9\\
     \cline{2-3}
    &Add content & 13\\
    \cline{2-3}
    &Remove content & 4\\
    \cline{2-3}
    &Change metadata & 7\\
    \hline
    
     \multirow{2}{*}{References}&Provide & 4\\
     \cline{2-3}
    &Request & 16\\
    \hline
     \multirow{2}{*}{Question}& On article's topic & 8\\
     \cline{2-3}
    & On Wikipedia conventions & 5\\
    \hline
	\multicolumn{2}{|c|}{Copyright issues} & 8\\
       \hline
       \multicolumn{2}{|c|}{Dispute claim in article} & 12\\
       \hline
       \multicolumn{2}{|c|}{General discussion about article's direction } & 8\\
       \hline
       \multicolumn{2}{|c|}{Other} & 6\\
       \hline

\end{tabular}
\caption{Categories of Wikipedia discussion page sections and number of examples in each category. Categories are based on a random sample of 100 discussion pages. Discussions can belong to more than one category.}
\label{Discussion_cat}
\end{table}

\xhdr{{Featured Articles}}
We now discuss a natural status measure for articles on Wikipedia.
The Wikipedia community chooses an article to feature every day through a \emph{peer review} process\footnote{See http://en.wikipedia.org/wiki/Wikipedia:Peer\_review for a description of Wikipedia's peer review process.}, and according to
their guidelines, such articles among
the very best in terms of 
professional standards of writing, presentation, and 
sourcing\footnote{See http://en.wikipedia.org/wiki/Wikipedia:Featured\_article\_criteria for a description of the attributes a featured article must have.}. 

We are interested in comparing coordination practices between highly
successful articles and average articles. 
While every measure of success includes particular idiosyncrasies
and potential biases, we believe that using the featured-article
designation as a success measure has a number of clear advantages.
In particular, rather than defining an ad hoc success measure ourselves,
the set of featured articles is a clear success measure that 
Wikipedia's own community has defined.
This has the advantage that our success measure is
likely to be compatible with the standards and goals of Wikipedia
editors, and it is something that produces incentives among editors.
It is certainly true that many very good
articles are never featured, but the existence of the designation
allows us to define a concrete and
very high standard of success for an article: whether it has been
featured or not.

\section{GitHub Data}
GitHub is a Git repository service used by millions of people to collaborate on open source software projects. Even though GitHub is smaller and more specialized than Wikipedia, it also exhibits the properties that make it a useful testbed for coordination in decentralized collaboration. We use data obtained from GitHub Archive\footnote{http://www.githubarchive.org/}, which provides a record of various aspects of all public repositories. From these data we are able to capture specific metrics of a project's visibility, size, and amount of coordination among collaborators. Our data contains all public projects that were actively developed during a three month period starting in May of 2012, which consists of about 300,000 projects. As with Wikipedia, we discuss how to develop measures of coordination 
and status for GitHub projects.

\xhdr{Coordination Mechanisms on GitHub}

When a user commits changes to a repository, others users have the option of
making comments or asking questions by issuing
a \emph{commit comment}. This feature allows collaborators to discuss
contributions and provide feedback. We use \emph{commit comments} to measure coordination on GitHub to understand how
people use comments on GitHub, we manually categorize a small
sample of comments. We follow the same methodology we use to
categorize Wikipedia comments. Table
\ref{Comment_cat_GitHub} shows the categories and the number of
examples in each category. We observe that commit comments are largely
used to discuss issues directly related to the project, and hence
serve as a reasonable measure of coordination.

\begin{table}[h!]
\centering
\begin{tabular}{ |c | c| }
  \hline
  \bf{Category}&  \bf{Num}\\
  \hline
  \hline
  Coding suggestion & 48\\
    \hline
  Code explanation by author & 3\\
    \hline
  Showing appreciation for other's work & 15 \\
    \hline
  Reporting bug & 13 \\
    \hline
    Question about other's code & 19 \\
    \hline
  General programming question & 4 \\
    \hline
  Expressing disapproval of other's work & 7 \\
    \hline
  Other & 15 \\
    \hline
\end{tabular}
\caption{Categories of GitHub comments and number of examples in each category. Categories are based on a random sample of 100 comments. Comments can belong to more than one category.}
\label{Comment_cat_GitHub}
\end{table}

\xhdr{GitHub Watchers}

In GitHub, users have the option of \emph{watching} repositories they are interested in. During the time period we are analyzing, watching a repository was a way bookmarking projects of interest\footnote{See https://developer.github.com/changes/2012-9-5-watcher-api/}. Since GitHub is mainly used to develop open source software and we only consider public projects, it is reasonable to assume that having many watchers signals high visibility. Hence, we use the number of watchers a project has as continuous measure of status. 

Having now articulated how the basic ingredients of our framework are
reflected in both the Wikipedia and GitHub, we turn to our
central questions in order.

\section{Coordination in High-Status Projects}

In this section we investigate how 
the amount of coordination varies with the status
of a project. For Wikipedia, we compare coordination in featured and
non-featured articles, and in GitHub we measure coordination as a
function of the project's number of watchers.

As noted in the introduction, we interpret our analysis via the
trade-off between the costs and benefits of explicit coordination 
\cite{Macmillan_04}:
While a highly coordinated team of collaborators has the advantage that it can split tasks, resolve disagreements, and set goals effectively, a team that works with little coordinating has the advantage that it spends all its efforts working on the task rather communicating with team members.

\xhdr{{The $x$-Core}}
It is known that in many on-line settings, a few users are responsible
for much of the content of the site \cite{Kittur_Kraut_2008,Kittur_Chi_Pendleton_Suh_Mytkowicz_2007}. 
We call this small group of users the \emph{core} of the project. More
concretely, we define the $x$-core of a project to be the smallest set of
users that account for an $x$ fraction of all the work;
these are the most active participants on the project.
As we increase $x$, the $x$-core gets larger as more participants
get included, and finally the $1$-core is the set of all
participants. We define the size of a project's $x$-core as the 
number of users it contains.  

The $x$-core can be defined for any collaborative project, but 
we focus here on its application to Wikipedia, because the 
projects there are large enough to show substantial variation as
we range over possible values of $x$.
By contrast, most of the GitHub projects we analyze are smaller and more
focused, and since the $x$-core analysis is correspondingly less
informative, we do not apply it in GitHub.

On Wikipedia,
a natural question is whether featured articles tend to have a larger
or smaller $x$-core size than non-featured articles. Measuring work by the number of edits, we compute the
fraction of editors that belong to the $x$-core of each featured article,
which captures the extent to which a small fraction of individuals
are doing most of the work.

\begin{figure}[htb!]
\centering
\includegraphics[width =.22 \textwidth]{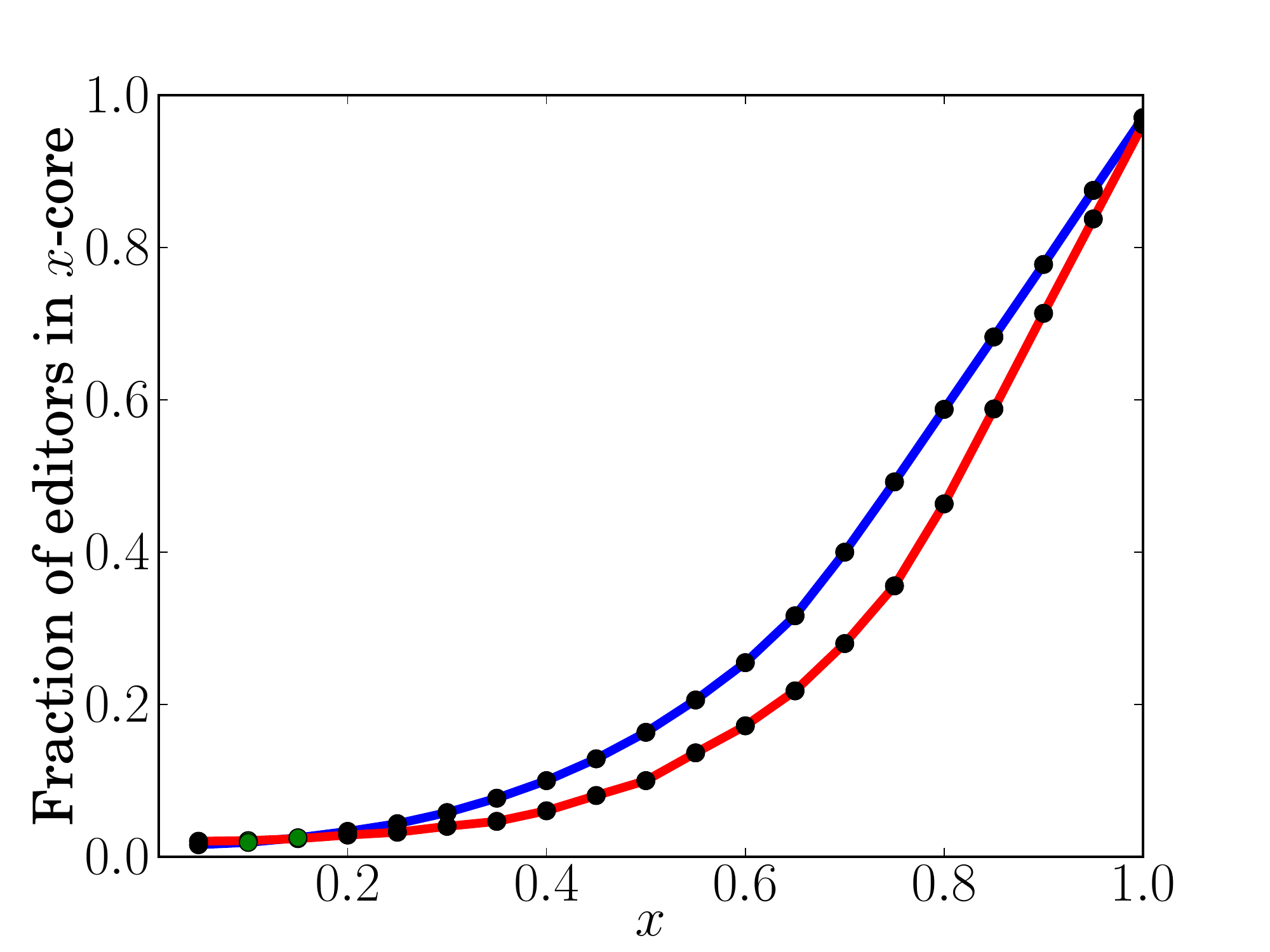} 
\caption{Median fraction of editors in $x$-core for featured articles (red) and non-featured articles (blue). \label{figures_coreSize}}
\end{figure}

Figure \ref{figures_coreSize} shows the median fraction of 
editors in the article's $x$-core for different values of $x$. Throughout the paper, statistical significance of the difference in medians using the Mann-Whitney U test \cite{Kruskal_1957} is indicated by the color of the dots: black (p-val $<$ .001), green (p-val $<$ .01), yellow (p-val $<$ .05). We
observe that featured articles have significantly smaller core sizes
for most values of $x$. 
Having few editors in the
core may be beneficial for an article
by making coordination among the core easier. 
In the same spirit as this result, Kittur and
Kraut found that Wikipedia articles benefit from having many editors
as long as a few editors are responsible for most of the edits
\cite{Kittur_Kraut_2008}. They propose that organizing in such a way
that a few editors are responsible for most edits is an implicit form
of coordination. 

\xhdr{Coordination in Featured Wikipedia Articles}
We now study how Wikipedia editors make use of explicit coordination
mechanisms to interact with each other, and how this operates
differently in featured and non-featured articles.
In this analysis, we consider the $x$-core for each $x$;
this lets us consider both the full article (when $x = 1$), as well
as whether coordination mechanisms are differentially used by the
most active editors (for smaller values of $x$).

\begin{figure*}[t]
\centering
\subfigure[Median number of discussion edits by $x$-core vs. $x$]{
\includegraphics[width =.22 \textwidth]{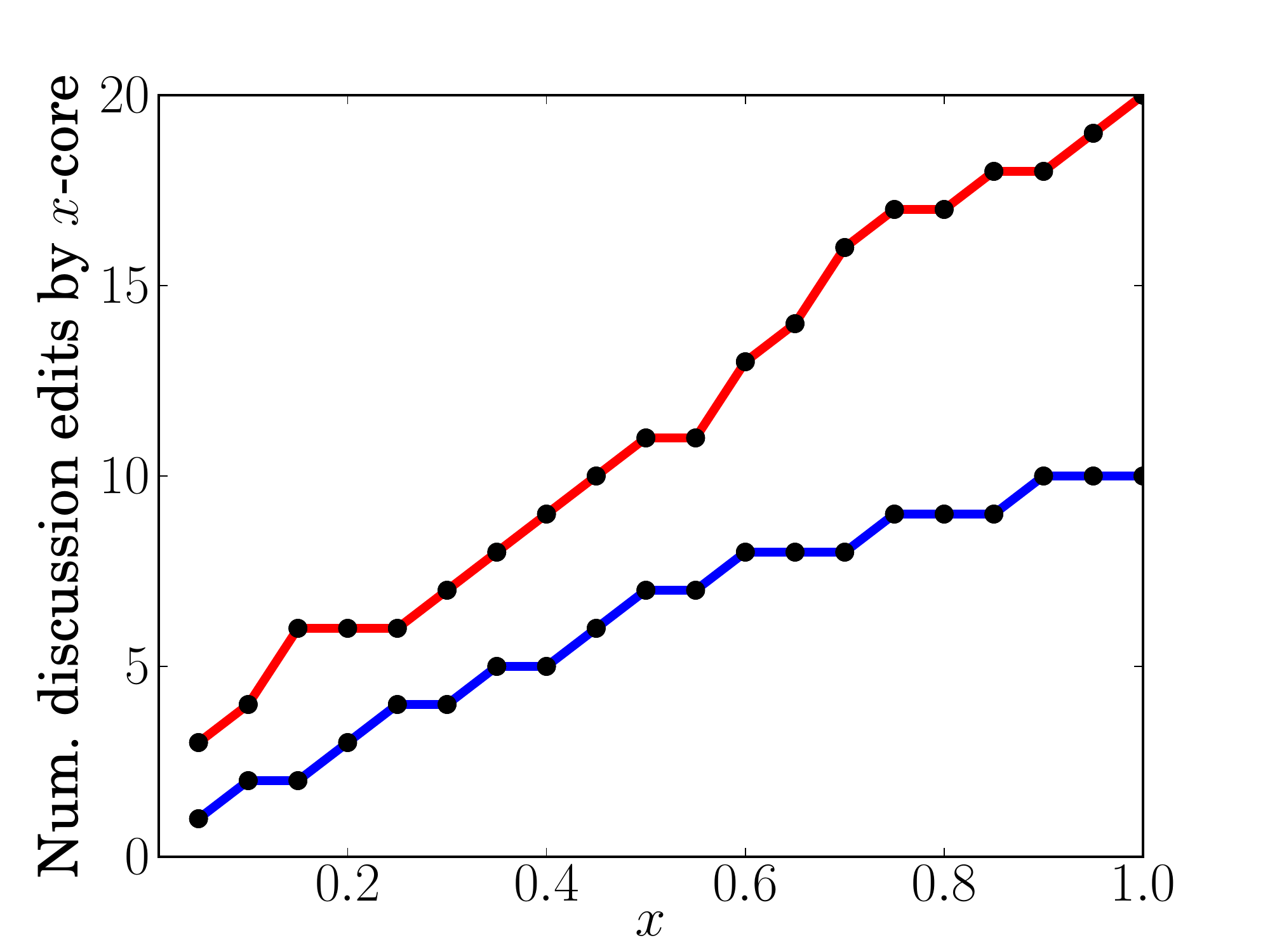} 
\label{DiscussionEditsNum_VS_XCore_MinEditors20_KSample_30}
}
\hspace{3mm}
\subfigure[Median value of $d(x) - x$ vs. $x$. $d(x)$ is the faction of discussion edits contributed by $x$-core editors.]{
\includegraphics[width =.22 \textwidth]{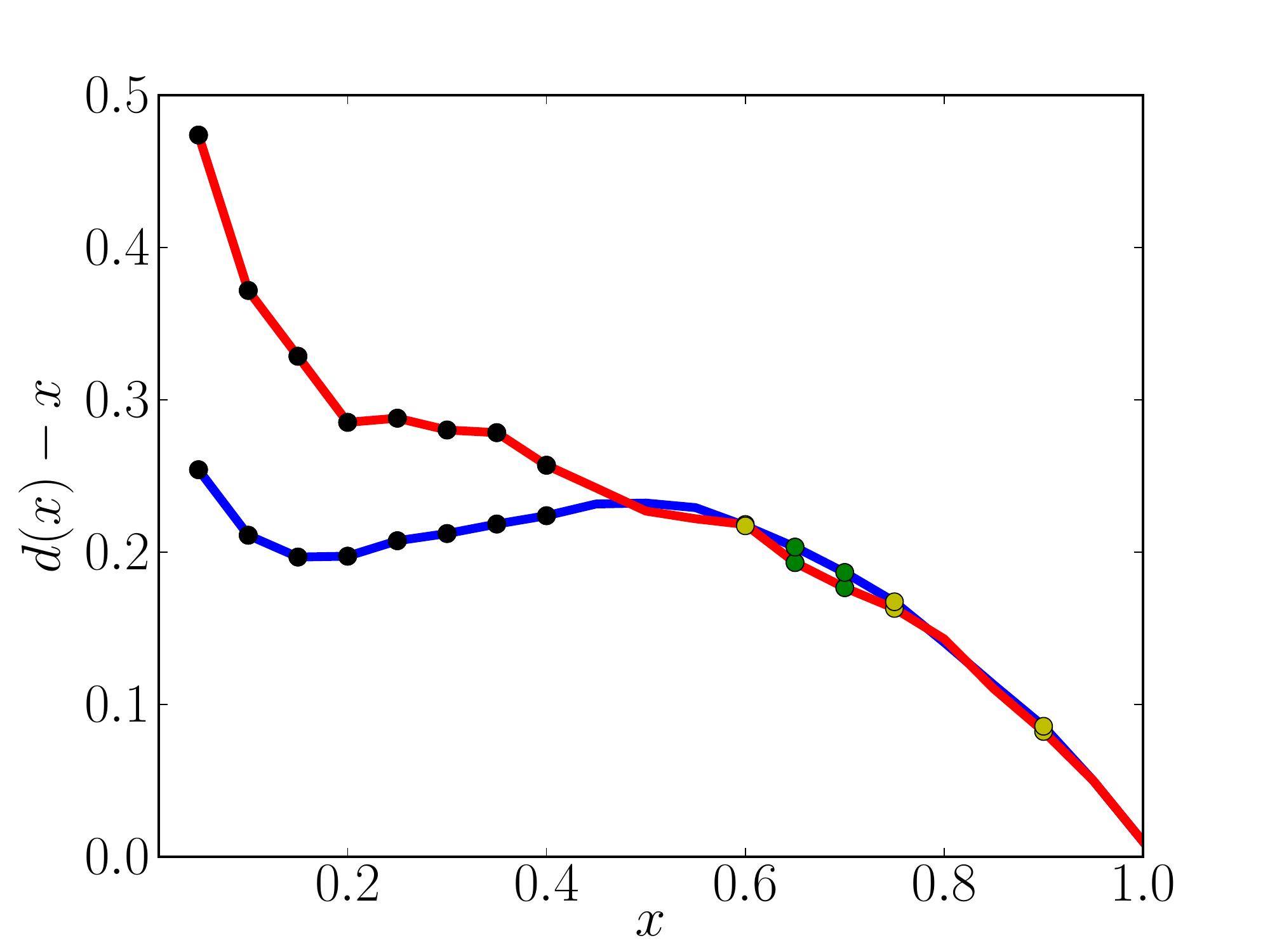} 
\label{DiscussionXCorePercentMinusXCore_Feat_VS_NonFeat_MinEditors_20_KSample_30}
}
\hspace{3mm}
\subfigure[Median number of comments by $x$-core vs. $x$]{
\includegraphics[width =.22 \textwidth]{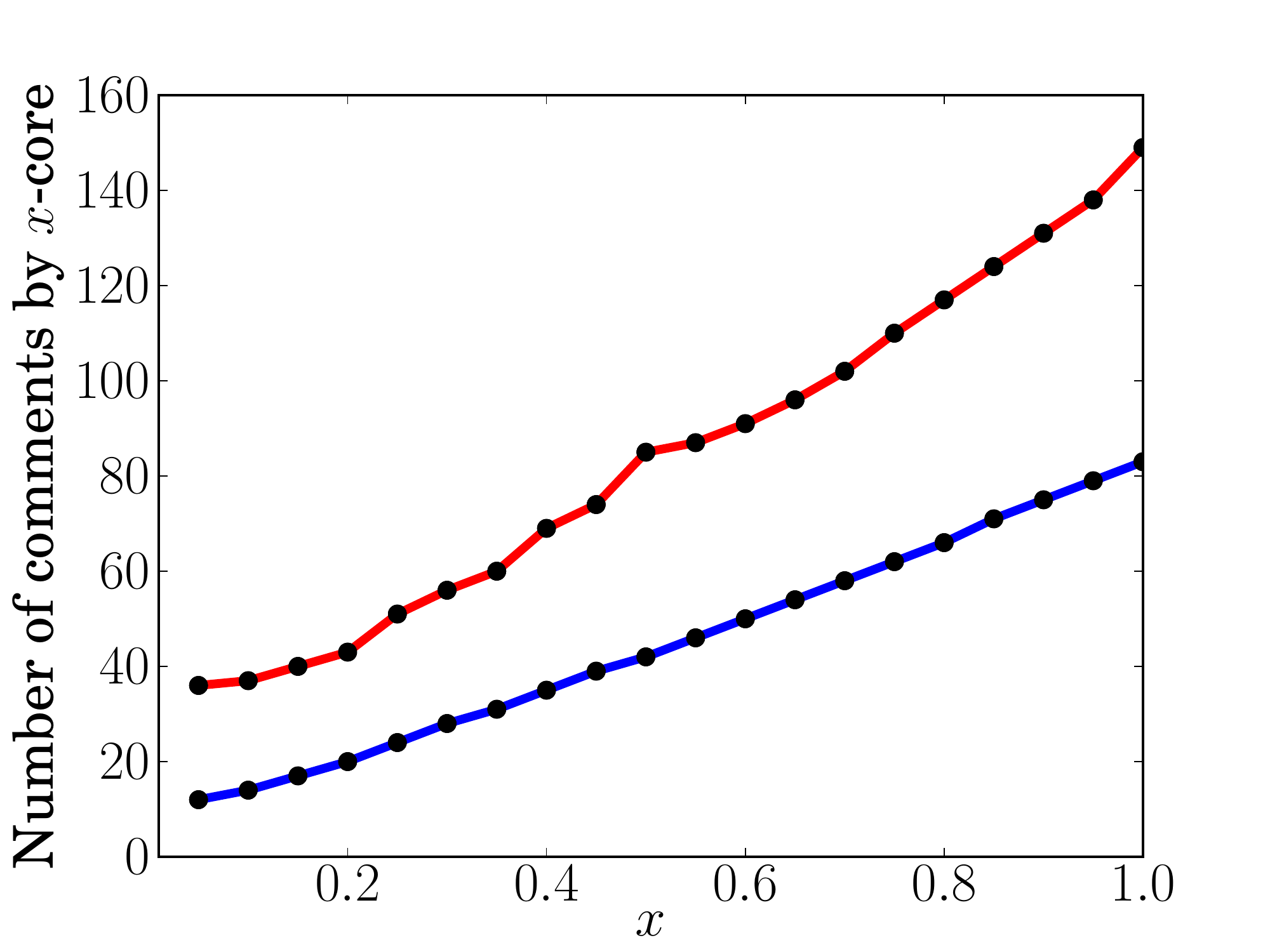} 
\label{CommentsNum_Feat_VS_NonFeat_MinEditors_20_KSample_30}
}
\hspace{3mm}
\subfigure[Median value of $c(x) - x$ vs. $x$. $c(x)$ is the faction of comments contributed by $x$-core editors.]{
\includegraphics[width =.22 \textwidth]{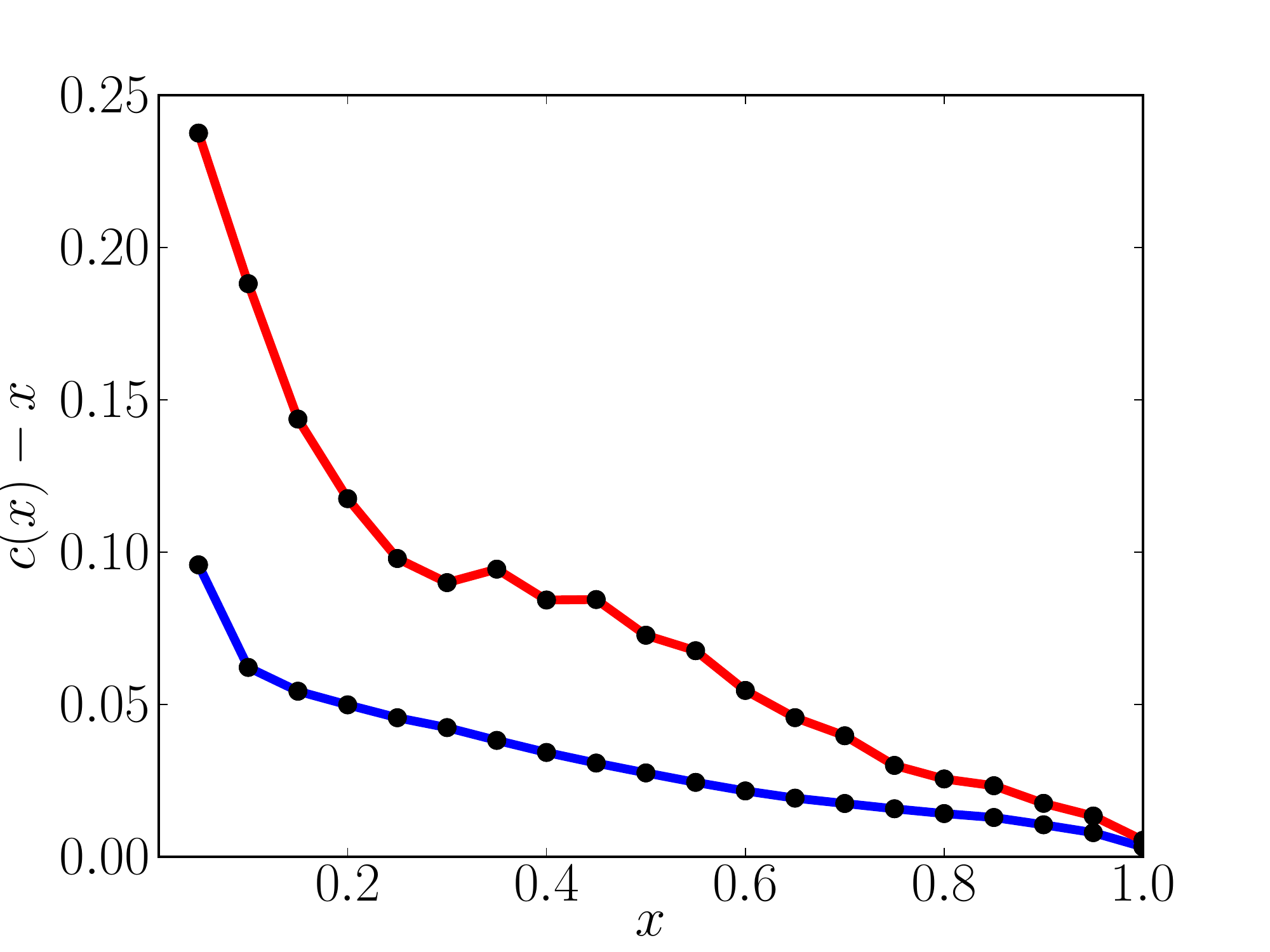} 
\label{CommentsXCorePercentMinusXCore_Feat_VS_NonFeat_MinEditors_20_KSample_30}
}
\caption{Differences in the use of comments and discussion pages among editors of featured (red) and non-featured (blue) articles. \label{figures_dicussionEdits_comments}}

\end{figure*}

We would like to compare featured and non-featured articles,
focusing on potential differences in coordination behavior.
We do this by constructing a comparison set --- a set of featured articles
and a set of non-featured articles over which we can 
compare aggregate properties.
Given the a priori differences between featured and non-featured articles,
such a comparison set needs to be constructed carefully so that
the contrasts we identify are not simply consequences of 
surface-level differences that we already understand.

In particular, we control for three different features in
setting up the comparison set.
First is the volume of edits: featured articles tend to have a
lot of editing activity before they are featured, and once they
are featured they attract even more attention than they
would get under normal circumstances.
Hence, when we compare editing behavior in featured 
and non-featured articles, we control for the number of edits. Second, we control for the stage of development of the article at
the time of its last recorded edit. Since articles on Wikipedia are
created every day, our data contains well established articles that
have been edited thousands of times as well as newer ones that have
only been edited tens of times. Finally, a third factor we control for is the 
stage of Wikipedia as a whole at the time 
when the edits to an article were done. Conventions among Wikipedia editors gradually
change over time, and the behavior of editors can be 
systematically affected by new conventions or features added on Wikipedia. 
Thus, our comparison set of featured and non-featured articles
comprises roughly the same distribution of time points from the
history of Wikipedia.

In summary, we construct two sets of articles with roughly the same distribution of edits that were generated at roughly the same time period --- one of featured articles and one of non-featured articles.
In the Appendix we provide the full details on exactly 
how this comparison set is constructed.

Let's now look separately at how discussion pages and comments
are used by featured and non-featured articles, considering the
$x$-core for multiple values of $x$. Given the coordination trade-off we discussed above, it is unclear whether featured articles should display higher or lower use of these coordination mechanisms.

Figures \ref{DiscussionEditsNum_VS_XCore_MinEditors20_KSample_30} and
\ref{CommentsNum_Feat_VS_NonFeat_MinEditors_20_KSample_30}
show that for all values of $x$, the $x$-core produces more 
discussion edits and comments in featured articles than non-featured articles.
Recall that our set of non-featured articles is constructed to
mirror the activity level of the featured articles as measured by
article edits, so the comparison in 
Figures \ref{DiscussionEditsNum_VS_XCore_MinEditors20_KSample_30} and
\ref{CommentsNum_Feat_VS_NonFeat_MinEditors_20_KSample_30}
is effectively saying that there is more coordination per edit in featured articles.

A distinct but related question is to consider which of the editors on
an article are accounting for the coordination activity.
In particular, there are two natural hypotheses: that the most
active editors are overrepresented in the discussion as they
coordinate; or, alternately, that the less active
editors are overrepresented in the discussion while their more active
counterparts do the work of writing the article itself.

To investigate this, we define $d(x)$ (respectively $c(x)$)
to be the fraction of
discussion-page edits (respectively comments) created by editors 
in the $x$-core, and we plot 
the differences $d(x) - x$ and $c(x) - x$ as functions of $x$.
Note that $d(1) - 1 = c(1) - 1 = 0$ by definition, and 
in the event that every editor contributed to the discussion pages and
comments in proportion to their article editing activity, 
we would have $d(x) - x = c(x) - x = 0$ for all $x$.

In Figures 
\ref{DiscussionXCorePercentMinusXCore_Feat_VS_NonFeat_MinEditors_20_KSample_30} and
\ref{CommentsXCorePercentMinusXCore_Feat_VS_NonFeat_MinEditors_20_KSample_30}
we show plots of $d(x) - x$ and $c(x) - x$ respectively, averaged 
separately over featured and non-featured articles.
The fact that these functions are positive over all $x < 1$ shows that
the more active editors (those who belong to the $x$-core for small $x$)
are in fact overrepresented in these coordination mechanisms.
Moreover, this overrepresentation is particularly pronounced for
the featured articles, again suggesting some of the distinctive
ways in which featured articles use coordination.
There is also an interesting contrast between discussion-page edits
and comments: $c(x) - x$ is higher for featured articles over all
$x < 1$, while $d(x) - x$ is higher for featured articles only for $x < 1/2$.
It would be interesting to further explore how this difference
relates to the lighter-weight nature of comments relative
to discussion-page edits.

\begin{figure}[t]
\centering
\subfigure[Median number of comments per commit vs. number of watchers]{
\includegraphics[width =.2 \textwidth]{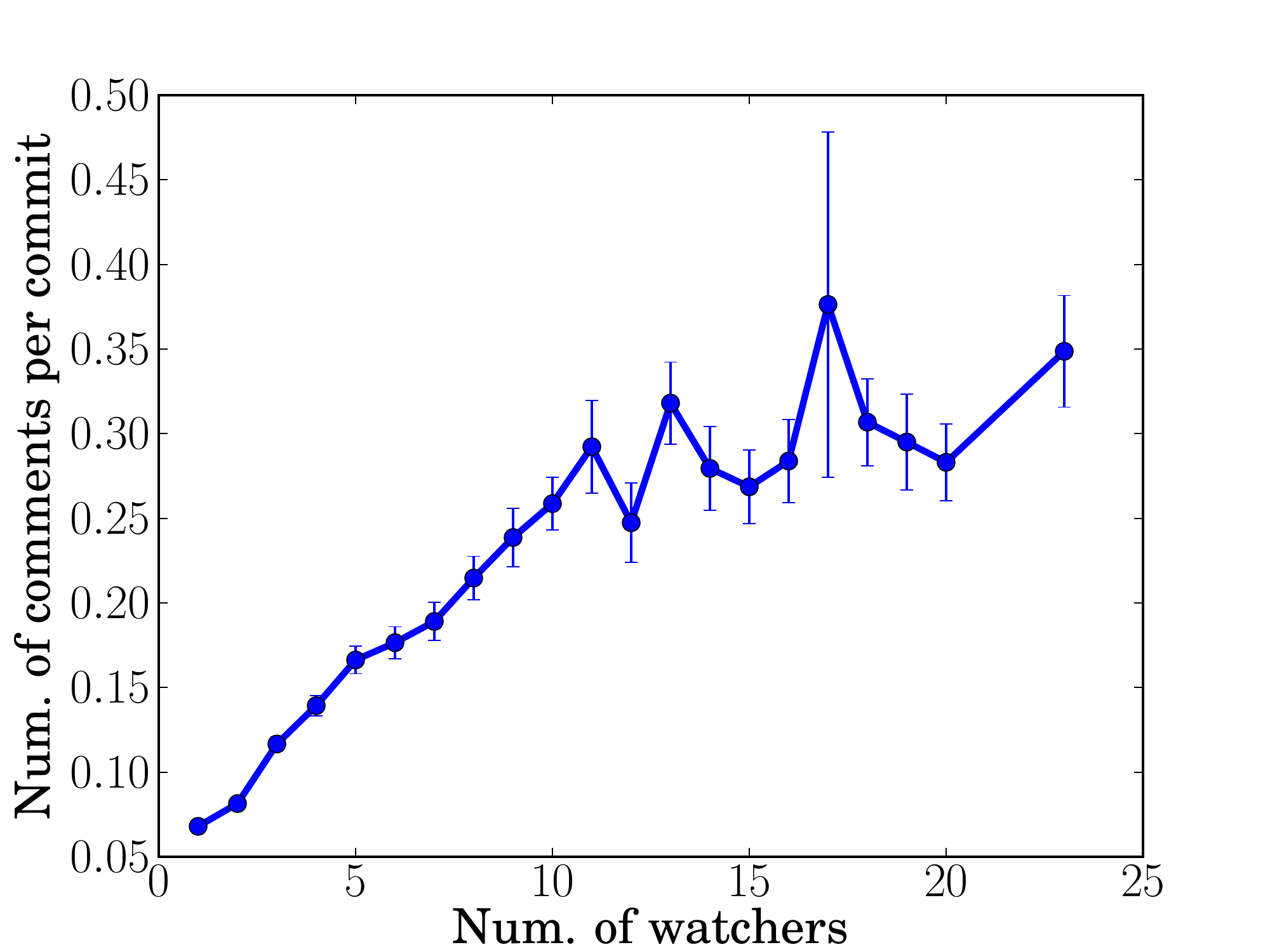} 
\label{wat_commentPerCommit_Min2Users}
}
\hspace{3mm}
\subfigure[Median number of comments per commit vs. number of commits.]{
\includegraphics[width =.2 \textwidth]{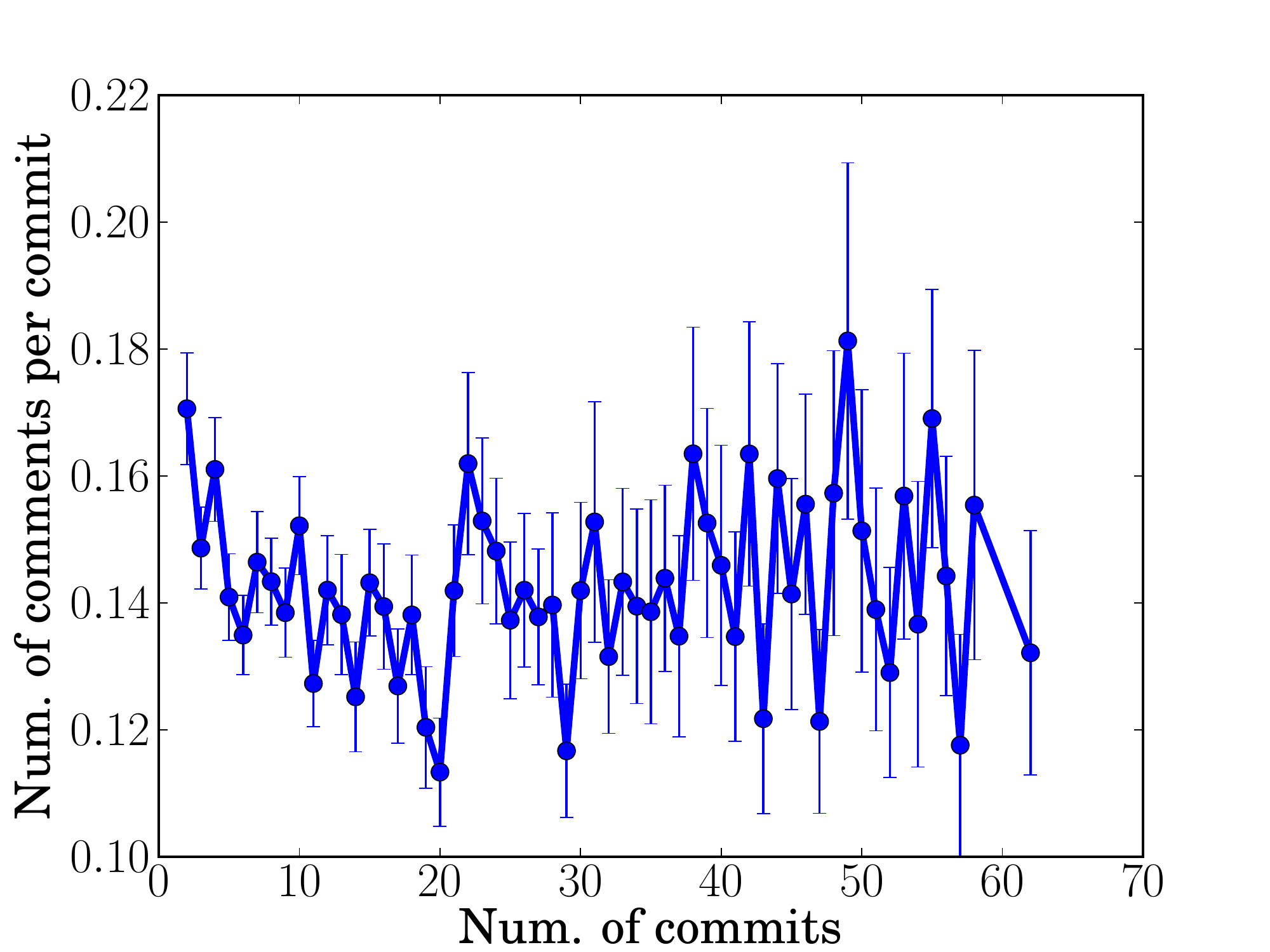} 
\label{commit_commentPerCommit_Min2Users}
}

\caption{Comments as a function
of the number of watchers (GitHub). 
\label{figures_dicussionEdits_comments}}

\end{figure}

\xhdr{Coordination in Highly-Watched Github Projects}
We now look at the relationship between coordination and status on
GitHub, keeping our discussion more brief for this dataset.
Since our measure of status in GitHub is continuous, rather
than comparing two sets of projects, we look at how the
number of comments per commit changes with number of watchers. Figure
\ref{wat_commentPerCommit_Min2Users} shows that the number of comment
per commit increases with number of watchers --- this trend 
points in the same direction as our Wikipedia analysis, with
higher-visibility projects using more coordination overall.

It is natural to ask whether the number of watchers is serving purely as a proxy for the number of commits, but as 
Figure \ref{commit_commentPerCommit_Min2Users} illustrates,
the number of comments per commit is roughly constant as a function
of the number of commits.  Thus the trend we are seeing
is not due to the number of commits, and this argues for the
relationship between visibility (number of watchers)
and the level of coordination.

We thus see that 
both GitHub projects and Wikipedia articles with higher status 
spend more effort coordinating. 
As noted earlier, this suggests the implication that projects
with higher visibility may be usefully guided in the direction of greater
levels of coordination relative to typical projects.

\section{Coordination in Crowded Environments}

Having established that projects with different status and visibility
in Wikipedia and GitHub can exhibit significant aggregate differences
in their use of coordination, we now explore the differences in
coordination among projects with different team composition. Our basic
intuition is that projects that involve a larger number of users
require more coordination than smaller teams. However, since our data
sets include projects that output different amounts of work, the size
of a team should be considered in relation to the amount of work it
produces. For a project with a fixed amount of work produced, we
expect that the amount of coordination will increase with team size.
Furthermore, for a project with a fixed team size, we expect that the
amount of coordination will decrease with amount of work produced. In
summary, our hypothesis is that more crowded projects (large team size
and low production) will require more coordination than less crowded
ones. This is because as projects become crowded by users, but the
amount of work available does not increase accordingly, users will
lose the ability to work separately on independent tasks, and more
communication will be necessary to coordinate multiple users working
on the same task. We now test our hypothesis on our data sets.

\xhdr{Crowdedness in Wikipedia}
We would now like to compare our hypothesis with what we
observe in Wikipedia. We first need to define measures
for the project's amount of work produced, number of users, and amount of
coordination. To define reasonable representations of these
parameters we take each Wikipedia article $a$ and consider the users
$U_a$ who have made at least one edit to the article and one edit to the discussion page. These are users who have demonstrated awareness
of the existence of the article's discussion page. 
Since articles in our data are at different stages, 
we consider a constant number of initial edits to measure the number of editors interested in the article and levels of coordination.  We record the timestamp
$T_a$ when the 100th edit by users in $U_a$ was made. We let
$S_a$ be the set of users in $U_a$ who contributed at least one
of the first 100 edits. The $S_a$ parameter represents the size of the team
as it existed at a fixed point in time.
We let $D_a$ be the number of discussion edits made before time $T_a$. 
The $D_a$ parameter represents the amount of coordination exhibited by the team. Here, we only consider discussion edits since they are the more explicit coordination mechanism. Finally, we let $N_a$ be the
eventual size of the article in bytes, which represents the amount of work produced by the editors. 

\begin{figure}[t]
\centering
\subfigure[Heat map of  ``amount of coordination" ($\log(D_a)$) as a function of the number of ``article parts" ($N_a$) and number of editors ($S_a$) in Wikipedia articles. ]{
\includegraphics[width =.2 \textwidth]{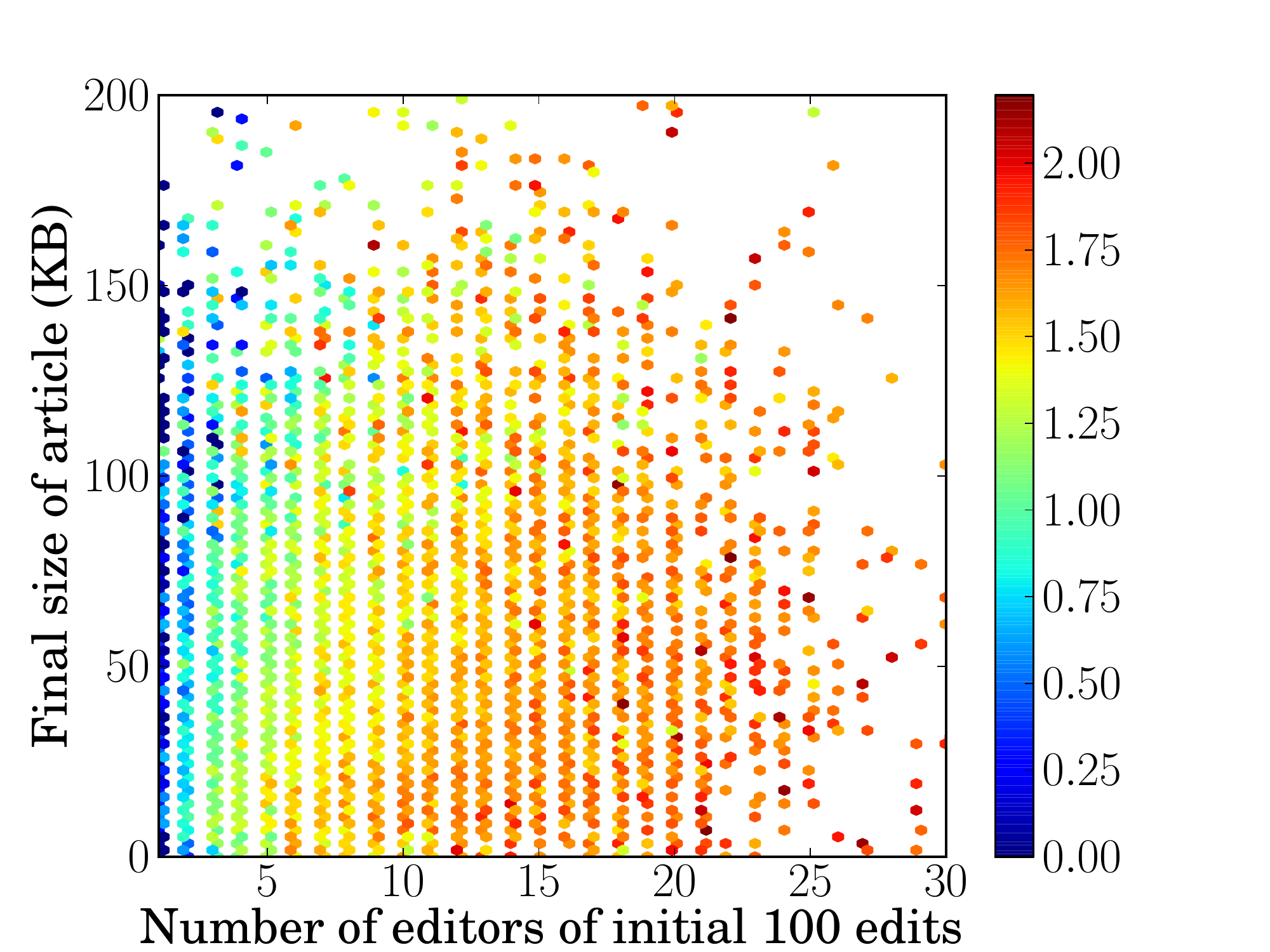} 
\label{Editors_VS_size_100_HeatMap}
}
\hspace{2mm}
\subfigure[Articles are split into 4 categories at the median size ($N_a$) and number of editors ($S_a$). Each area shows the median number of discussion edits ($D_a$), median number of discussion edits per editor ($D_a/S_a$), and number of articles ($N$). Differences between cells are statistically different (p-val $<0.05$)]{
\includegraphics[width =.2 \textwidth]{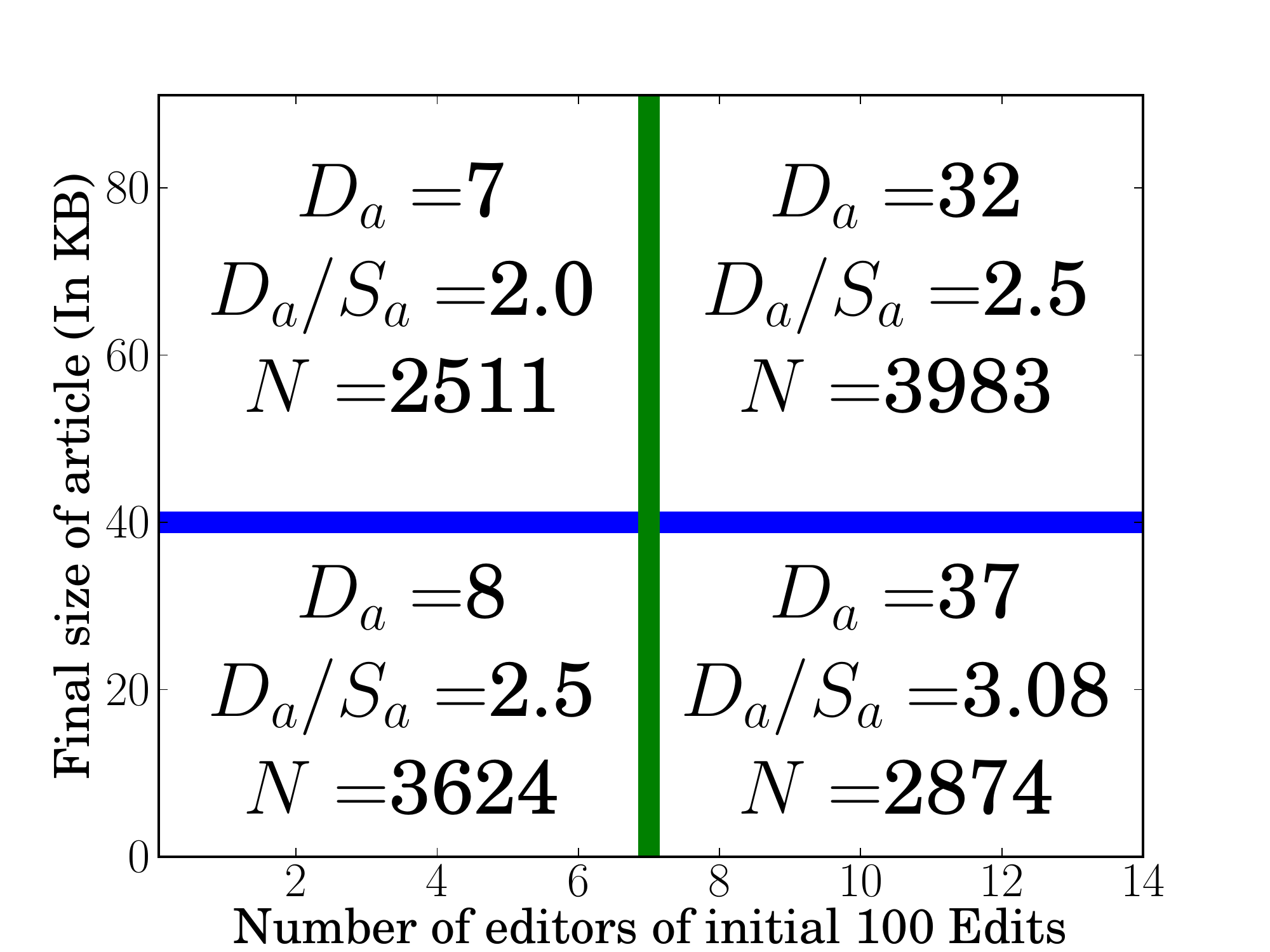} 
\label{Size_VS_Editors_4Categories}
}
\caption{Amount of coordination in Wikipedia articles of different composition. \label{GitHub_heatmap}}
\end{figure}

Figure \ref{Editors_VS_size_100_HeatMap} depicts the amount of coordination $\log(D_a)$ in an article of size $N_a$ and $U_a$ editors. It is drawn as a heat map, with the color corresponding to the value of $\log(D_a)$. Furthermore, in figure \ref{Size_VS_Editors_4Categories}, we split articles into four categories depending on whether ---
relative to the median article --- they have
a lower or higher number of bytes ($N_a$) and a 
lower or higher number of editors ($S_a$). 
We compare the number of
discussion edits ($D_a$) and the number of discussion edits per editor
($\frac{D_a}{S_a}$) among the four categories. 

Figures \ref{Editors_VS_size_100_HeatMap} and \ref{Size_VS_Editors_4Categories} show the general trend we hypothesized. The amount of coordination, measured by the number of discussion edits and discussion edits per editor, increases with the number of editors and decreases with the size of the article. Articles that are crowded with many users and low production exhibit the most coordination. 

\xhdr{Crowdedness in GitHub}
We now explore the role of crowdedness in the coordination of GitHub projects. 
We measure amount of work produced, number of users, and amount of
coordination analogously to how we measured them for Wikipedia,
and then perform an analogous investigation of the relationship.
For each GitHub project $p$ we let $U_p$ be the users who committed at least once and contributed at least one comment. We record the timestamp
$T_p$ when the 100th commit by users in $U_p$ was made. We let
$S_p$ be the set of users in $U_p$ who contributed at least one
of the first 100 commits. We let $C_p$ be the number of
comments made before $T_p$. Finally, we let $N_p$ be the
eventual size of the project measured by 
the total number of commits in 
the full history of the project. We use $N_p$ as the amount of work produced, 
$S_p$ as the number of users, and
$\log(C_p)$ as the ``amount of coordination.''

\begin{figure}[t]
\centering
\subfigure[Heat map of  ``amount of coordination" ($\log(C_p)$) as a function of percentile of number of ``project parts" ($N_p$) and number of users ($S_p$) in GitHub projects]{
\includegraphics[width =.2 \textwidth]{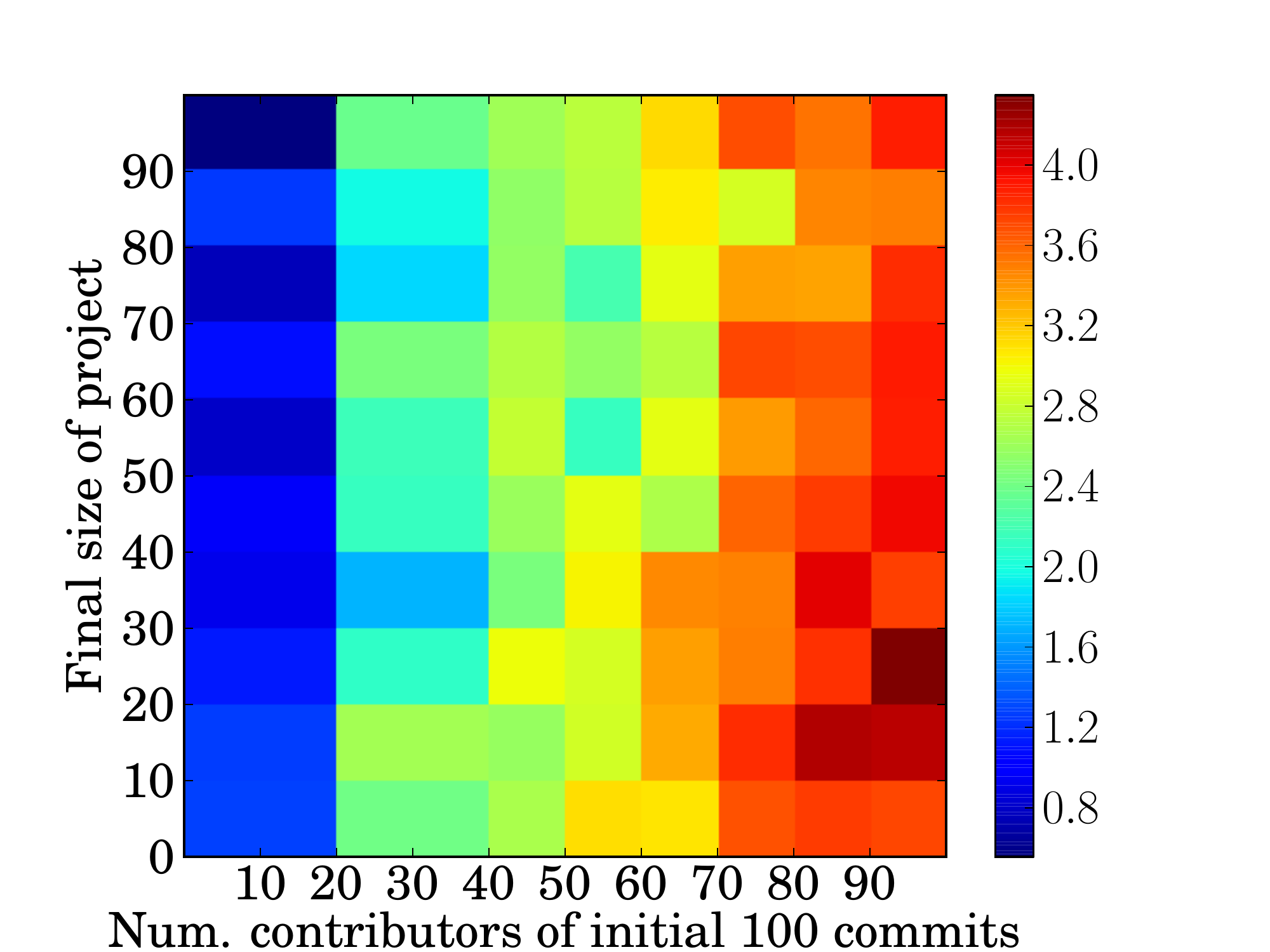} 
\label{contributors_vs_commits_percentiles_firstK_100}
}
\hspace{1mm}
\subfigure[Split of projects into 4 categories by low or high size ($N_p$) and number of editors ($S_p$). Each area shows the median number of comments ($D_p$), median number of comments per user ($D_p/S_p$), and number of projects ($N$). Differences between cells are statistically different (p-val $<0.05$)]{
\includegraphics[width =.2 \textwidth]{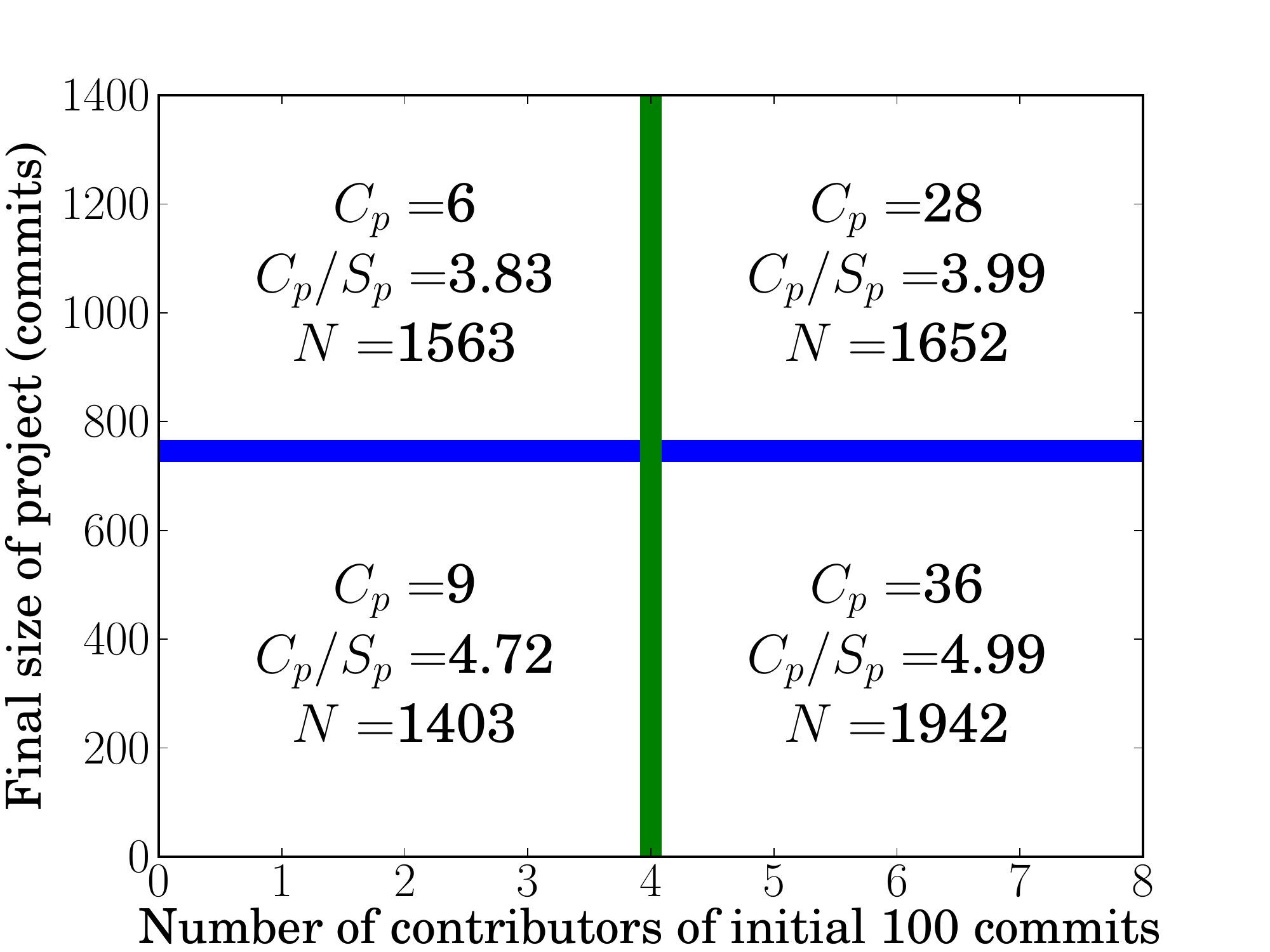} 
\label{contributors_size_split_at_median}
}
\caption{Amount of coordination in GitHub articles of different composition. \label{GitHub_heatmap}}
\end{figure}

Since the GitHub data set is much smaller than Wikipedia, in
order to see the change in coordination with project size and number
of users, we split the projects into 100 bins by splitting the number
of users ($S_p$) and the size of the projects ($N_p$) into 10 bins by
percentile. We then measure the amount of coordination ($\log(C_p)$)
within each bin. Figure
\ref{contributors_vs_commits_percentiles_firstK_100} shows a heat map
of coordination $\log(C_p)$ as a function of number of users ($S_p$)
and the size of the projects ($N_p$). To observe a numerical
representation of the trend, we also split the projects into four
categories depending on whether they have low or high number of users
and project size, relative to the median. Figure
\ref{contributors_size_split_at_median} shows the median values of
$D_p$, $\frac{D_p}{S_p}$, and the number of projects in each category.

We find that the trend is similar to the one observed in Wikipedia. As projects become more crowded with many users and small size, users tend to coordinate more. 
Overall, the fact that coordination and crowdedness align closely in both domains raises a further potential
implication for design --- to recognize projects that are
becoming increasingly crowded, and to correspondingly guide groups toward
coordination resources as this is occurring.

\subsection{Modeling Coordination and Crowdednes}

We now develop a simple theoretical model of coordination, 
so that we can study the cost-benefit trade-off in coordination
at a more formal level.
The model is highly stylized to reduce this trade-off
to its basic essence. The benefit of working with a concrete stylized
model is that by stripping away much of the complexity of the
coordination process that is specific to Wikipedia and GitHub, we can
highlight how certain trade-offs depend only on
a set of simple assumptions that are present in other
domains outside of Wikipedia and GitHub. We find that the model
accomplishes this -- from the analysis of the model, we identify the
relationship between crowdedness and coordination, which matches our
original hypothesis and the effects we found in Wikipedia and GitHub.

\xhdr{General Setting for the Model}

We begin by modeling the costs and benefits of coordination
in a stylized setting that represents a team's collaboration.
The costs result from the fact that time spent on coordination
does not directly advance the project itself.
The benefits will be based on dividing
up the work without redundancy --- in the absence of 
coordination, there is the danger
that two people will try to do the same step in the project
simultaneously, resulting in a loss in efficiency.
In the context of GitHub and Wikipedia, we can think of this model as
capturing the way in which 
coordination is particularly important when two users are
working on the same section of an article or the same part of a program. 
Without coordination or
discussion, users who disagree on the outcome 
often engage in ``edit wars,'' undoing
each other's work \cite{Viegas_2004}.

As discussed above, 
our model is not designed to capture all the nuances encountered
in contexts as complex as Wikipedia and GitHub. Indeed, our goal is in 
a sense the opposite, to find the simplest formulation of a model
in which the inherent trade-off between coordination and efficiency
emerges from the basic properties of the model.

The structure of our model is as follows. There is a project with $N$ {\em parts} that
need to be finished. Each part starts in the 
{\em unfinished} state, and requires one unit of work;
a part transitions from the unfinished to the finished state
when a user works on it.  A set of $E$ users work on the project.
In a single step, a user contributes a unit of work to one part.
If the part is currently unfinished, then it becomes finished.
If the part the user worked on was already in the finished state,
the situation is more subtle, reflecting the collision between two users.
The new contribution has no effect with probability $1 - \alpha$;
and with probability $\alpha$ it in fact has a negative effect,
clashing with the previous contribution and returning it to
the unfinished state.  This captures the idea noted above, that
when two people work on the same task, it can create additional work as the 
differences are resolved.

Users arrive sequentially and can either
coordinate with previous users before contributing,
or contribute without coordinating. 
Specifically, each user is allowed to perform two actions, selected in one of the following two ways:
\begin{itemize}
\item {\em Coordinate:} If the user coordinates,
she uses her first action to coordinate with others 
to find an empty project part, and then uses her second action 
to add a contribution to the empty part, moving it to the finished state.
In this case, she contributes
exactly one part to the project. 
\item {\em Not Coordinate:}  If the user does not
coordinate, then she uses 
both of her actions to sequentially contribute to two 
randomly selected project parts. Either of these parts might
turn out to be finished, in which case the contribution has no effect
(with probability $1 - \alpha$) or returns the part in question
to the unfinished state 
with probability $\alpha$. 
\end{itemize}

Each user chooses to coordinate independently 
with probability $\beta$ and to not
coordinate with probability $1-\beta$, where $\beta$ is fixed in the
beginning of the process. The two actions allowed to users in the model represent the ability that users in decentralized collaboration platforms have to spend all the effort contributing to the projects, or to split their efforts between contributing and coordinating. For example, as we observed, some editors justify their edits in the discussion page and some propose their edits before applying them to the article. Likewise, some GitHub users explain their code after committing it.

Collectively, the set of users would like 
the project to have as many finished parts as possible. How much should they coordinate in order to maximize this objective? We
are interested in finding the value of $\beta$ that maximizes the
number of finished parts by the end of the process.

\begin{figure}[t]
\centering
\subfigure[Values obtained from simulations. Parameter $\alpha = 1$.]{
\includegraphics[width =.22 \textwidth]{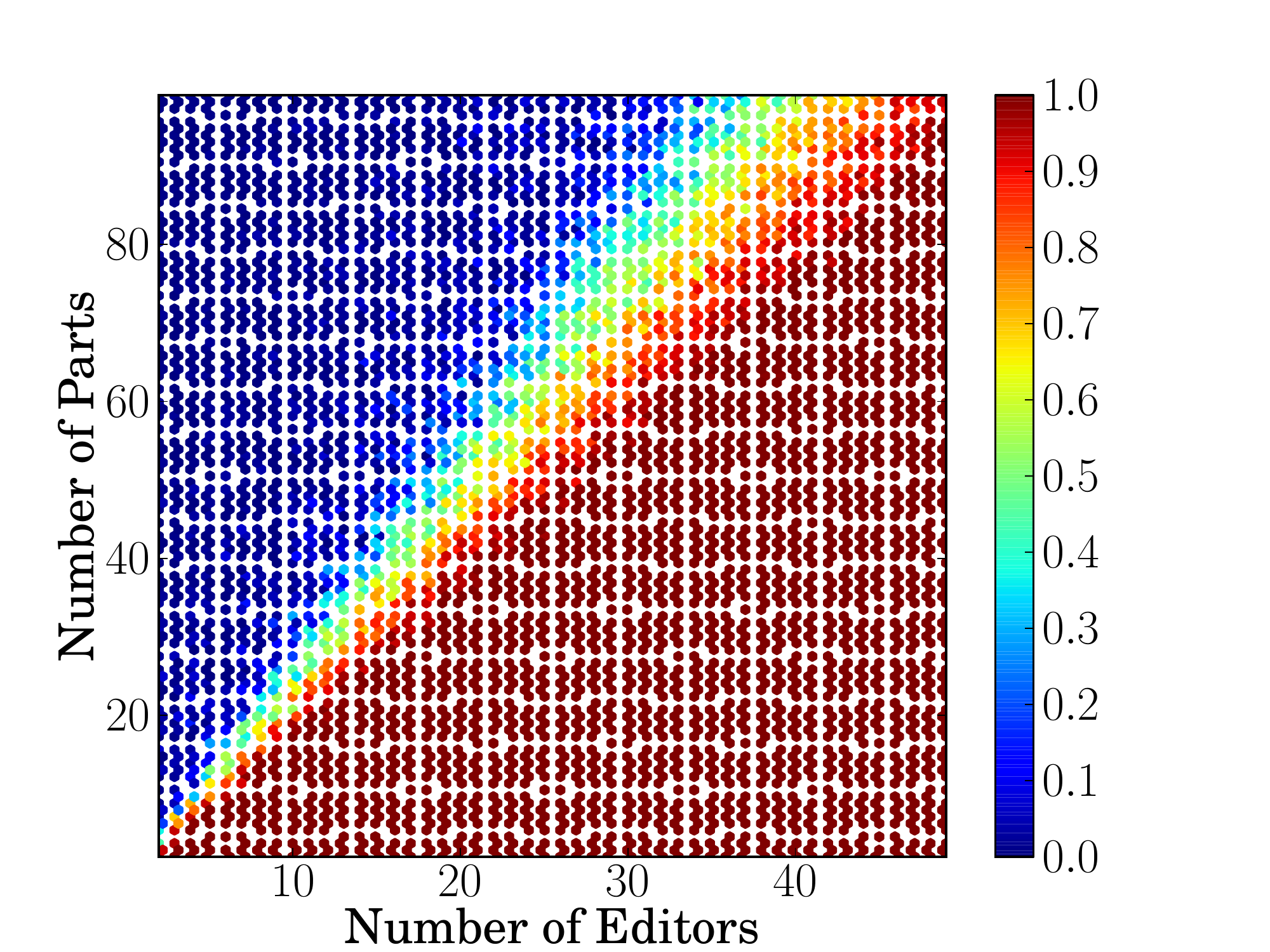} 
\label{Heatmap_N_alpha_100_50_colorbar_PenaltyRandomlySelected0Or1_Prob50}
}
\subfigure[Values obtained from analytical approximation. Parameter $\alpha = 1$.]{
\includegraphics[width =.22 \textwidth]{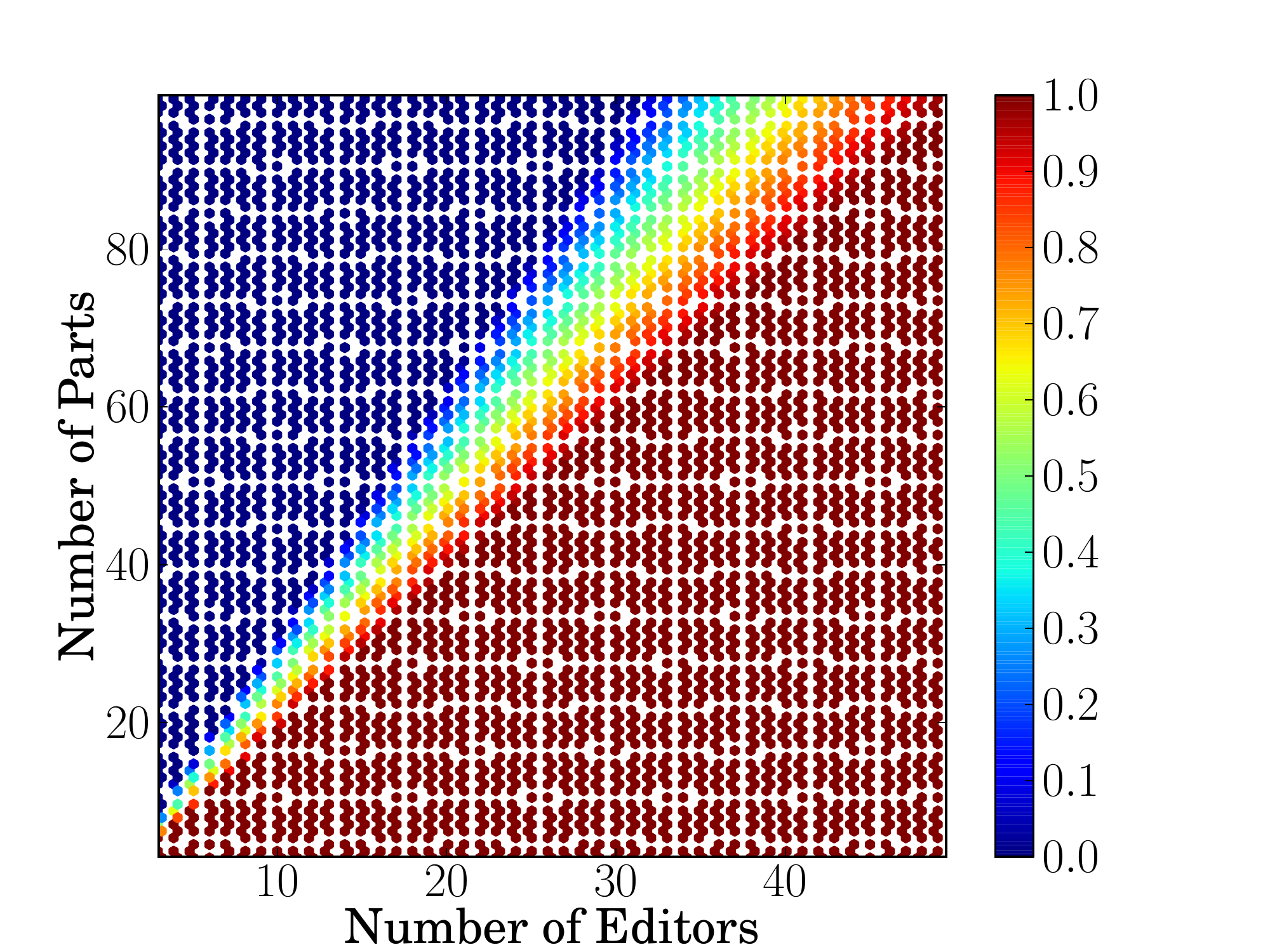} 
\label{Editors_VS_size_100_HeatMap_AnalyticalFromSolvingRecurrence.pdf}
}
\subfigure[Values obtained from analytical approximation. Parameter $\alpha = 0$.]{
\includegraphics[width =.22 \textwidth]{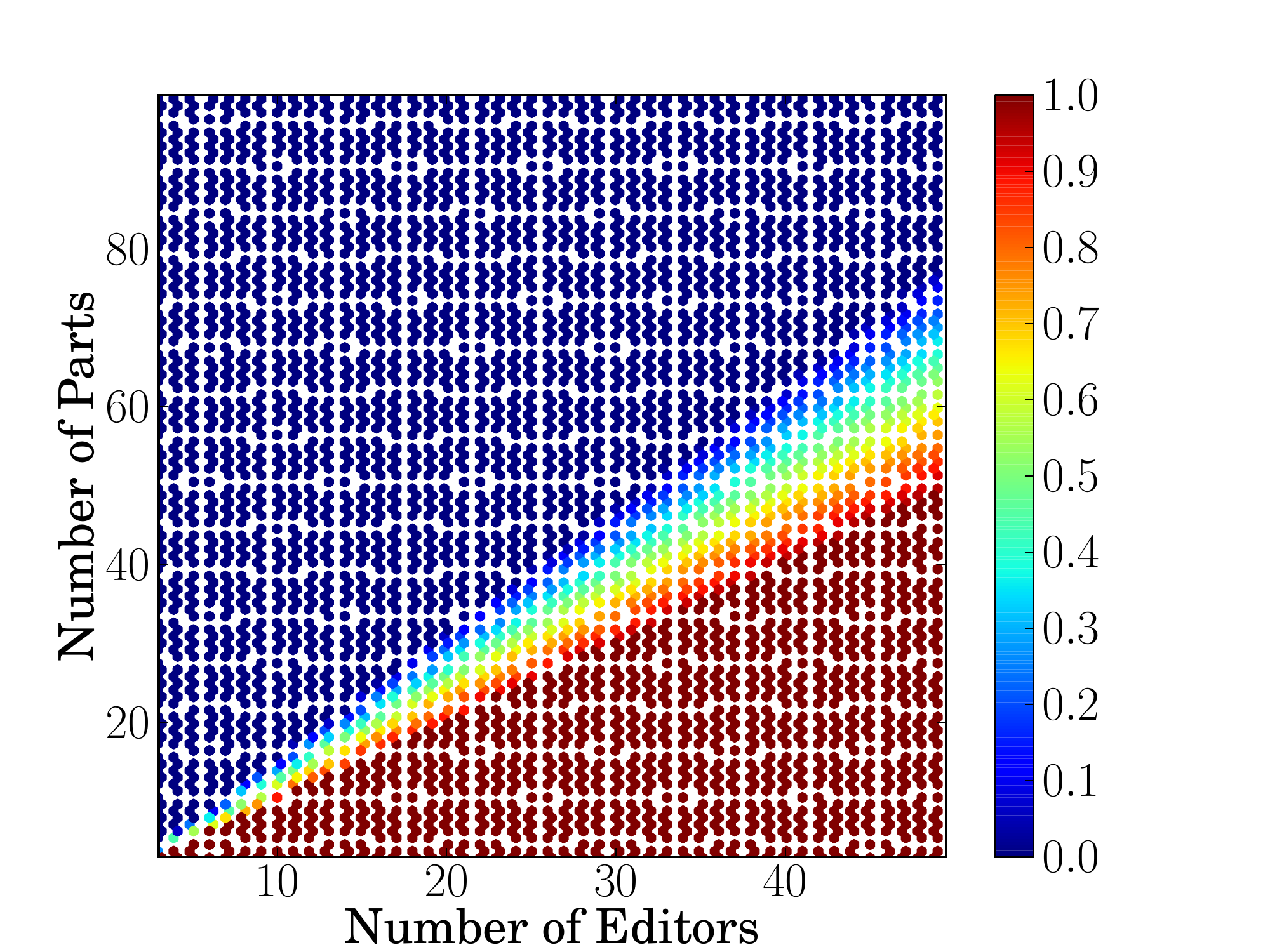} 
\label{Editors_VS_size_100_HeatMap_Alpha0_AnalyticalFromSolvingRecurrence}
}
\caption{Predicted amount of coordination by the model. Heat maps show the amount of coordination ($\beta$) that optimizes the number of finished parts of a project with $E$ workers and $N$ parts. \label{simultaneous_simulations_and_data}}

\end{figure}

\omt{
\begin{table}[t]
\centering
\begin{tabular}{ c | c }
  $\displaystyle k$ &  $\displaystyle X_{C,k}$\\
  \hline
  \hline
  2 & $\displaystyle \frac{(N-C)(N-C-1)}{N^2}$\\
  1 & $\displaystyle(1-\alpha)\left[\frac{C(N-C)}{N^2}+ \frac{(N-C)(C+1)}{N^2}\right]$\\
  -1 & $\displaystyle \alpha(1-\alpha)\left[\frac{C(C-1)}{N^2} + \frac{C^2}{N^2}\right]$ \\
  -2 & $\displaystyle\alpha^2\left[\frac{C(C-1)}{N^2}\right]$ \\
\end{tabular}
\caption{Probability that user who does not coordinate
finishes $k$ parts given that there are $C$ finished
parts at the time the user arrives}
\label{Values of X}
\end{table}
}

Figure \ref{Heatmap_N_alpha_100_50_colorbar_PenaltyRandomlySelected0Or1_Prob50} depicts the optimal
$\beta$ with $N$ parts, $E$ users, and with $\alpha = 1$.
It is drawn as a heat map, with the color corresponding to the
value of $\beta$. Here $\beta$ has been optimized 
by running a large number of simulations of the process. 
We observe first of all from the heat map that the relationship 
between crowdedness and coordination in this synthetic model
follows the same qualitative direction that we observed for
both Wikipedia and Github.
We now turn to an analytical approximation of the optimal $\beta$ 
and show it agrees very closely with the 
simulation results.

\paragraph*{\bf Analysis of optimal coordination probability}
Fix $N$, $E$, and $\alpha$, and let $P_i$ be the expected number of
finished parts the project has after $i$ users have passed through the project. Also, let $X_{C, k}$ be the probability that a user who does not coordinate
finishes $k$ parts given that there are $C$ finished
parts at the time the user arrives. When $k< 0$, $X_{C, k}$ is the
probability that the user will \emph{undo} $k$ parts from the
project. We make the following approximation:
\begin{equation}
P_{i+1} = (1-\beta)(P_i +  2X_{P_i,2} + X_{P_i,1} - X_{P_i,-1}-2X_{P_i,-2}) + \beta
\label{Recurrence_exp_val}
\end{equation}
Note that this is an approximation since the exact value of $P_{i+1}$
is $(1-\beta)(C_i +  2X_{C_i,2} + X_{C_i,1} - X_{C_i,-1}-2X_{C_i,-2})
+ \beta$, where $C_i$ is the actual number of finished parts after
user $i$, not the expected value of $C_i$. That is,
we make the deterministic approximation
that $C_i$ is always its expected value.

Writing the values of $X_{C,k}$ in terms of $\alpha$, $C$, and $N$
and plugging them into equation \ref{Recurrence_exp_val},
we get the following recurrence relation:
\begin{equation}
\begin{aligned}
P_{i+1} &= AP_i + P_0\\
\text{where } A &= \frac{1-\beta}{N^2}(1+\alpha)^2  - 2\frac{1-\beta}{N}(1+\alpha) + 1\\
 \text{and } P_0 &= \displaystyle -\frac{1-\beta}{N}(1+\alpha) + 2 - \beta\\
\end{aligned}
\label{recurrence_simplified}
\end{equation}

Solving the recurrence, we get a closed form solution:

\begin{equation}
\begin{aligned}
P_i &= P_0\left[\displaystyle\frac{A^i-1}{A-1}\right] \text{ if } A \ne 1\\
\text{and } P_i &= iP_0 \text{ if } A = 1
\end{aligned}
\label{closed_form}
\end{equation}

For fixed $N$, $\alpha$, and $E$, $P_{E}$ in equation \ref{closed_form} gives the expected number of completed parts as a univariate function of $\beta$. We can easily optimize this function and solve for the value of $\beta$ that leads to the most completed parts. Figure \ref{Editors_VS_size_100_HeatMap_AnalyticalFromSolvingRecurrence.pdf} shows the optimal value of $\beta$ for a range of values of $N$ and $E$ and $\alpha = 1$. We observe that the result from the simulations (figure \ref{Heatmap_N_alpha_100_50_colorbar_PenaltyRandomlySelected0Or1_Prob50}) and the analytical approximation (\ref{Editors_VS_size_100_HeatMap_AnalyticalFromSolvingRecurrence.pdf}) are very similar, suggesting that both approaches lead to a good approximation of the optimal solution.  

We notice first that when $E > N$ the best $\beta$ is 1,
for the simple reason that in this case, if all users coordinate, they will finish all the parts -- the best possible outcome. Conversely, when $N$ is very large and $E$ is small
then the best $\beta$ is close to zero. That is because there are many more
parts than users, so it is unlikely for users to collide, and hence
coordinating is mainly wasteful. In between these two extremes, there are
values of $N$ and $E$ in which the best value for $\beta$ 
lies non-trivially away from both zero and one.

In the previous analysis we set $\alpha = 1$. We now show that the high level predictions of the model hold for any value of $\alpha$. Figure
\ref{Editors_VS_size_100_HeatMap_Alpha0_AnalyticalFromSolvingRecurrence}
shows the optimal $\beta$ for a range of value of $N$ and
$E$ using $\alpha = 0$. 
We observe that
the area where the optimal value of $\beta$ is strictly between 0 and 1
rotates clockwise and there is overall less coordination. However, the basic
trend observed when $\alpha = 1$ holds here too ---
projects require more coordination when the number of
project parts is smaller relative to the number of users. Analogous plots across various values of $\alpha$ between 0 and 1 exhibit the same general trends. 
For all $\alpha$
the model thus makes a basic qualitative prediction:
users should coordinate more as the project becomes more
``crowded,'' with $E$ large relative to $N$. The alignment of this stylized model with our hypothesis and the trends we observed in Wikipedia and GitHub suggests the potential robustness of the relationship between crowdedness and coordination in other collaborative domains. It is striking that this relationship emerges clearly from the model, despite the fact that crowdedness was not explicitly built into the model's structure.

\section{Discussion}

Through the use of rich datasets 
and a theoretical model of coordination, 
we have analyzed the performance of on-line projects 
from the perspective of coordination mechanisms.
On both Wikipedia and GitHub,
we find that projects with high status and visibility
differ in aggregate from
other projects in the way that they use coordination. We also find that crowding of project participants
is a key parameter underlying the coordination level in Wikipedia and GitHub. We develop a theoretical model for the coordination process;
the analysis of the model aligns with the trends found in the data, suggesting the potential robustness of the findings.

The relationship between coordination and these structural properties
of projects can suggest principles for designing coordination mechanisms
in several dimensions.
\begin{itemize}
\item The relationship between crowdedness and coordination suggests
that coordination mechanisms should not be surfaced uniformly across
different projects, but instead emphasized more strongly on \emph{crowded} projects -- those with many team members relative to the project's size.
 
\item In a related vein, our analysis has pointed to differences
between lightweight and heavyweight coordination mechanisms, especially
in their differential usage across active and peripheral team members.
There is thus a need to integrate these different coordination styles
across different types of contributors.

\item Featuring and visibility interact
in subtle ways with coordination, as we have seen; there may also be
additional dimensions along which coordination should most effectively
be varied.
\end{itemize}
Broadly speaking, our framework here suggests that an understanding of
the roles of coordination mechanisms in different settings 
can benefit from a data-oriented analysis of their inherent trade-offs.

The similarities in these effects across both Wikipedia and GitHub
suggests some of their generality; it will be interesting to consider
how these findings carry over to other on-line domains in which 
decentralized teams collaborate on projects.
As we have seen from the two domains in this paper,
the basic ingredients of our model and analysis can usually be directly
adapted to new settings, since the framework can be applied
whenever there is a group faced with a primary work product and
a separate channel for exchanging coordination-related messages.
Ultimately, seeing how these findings transfer across domains
will be a next useful step on the way toward understanding the
process of large-scale on-line collaboration.

 \xhdr{Acknowledgments}
 
 This work was supported in part by a Simons Investigator Award, a Google Research Grant, a Facebook Faculty Research Grant, an ARO MURI grant, and NSF grant IIS-0910664

\section{Appendix: Sampling Wikipedia Articles}
\label{materials}

In this appendix we describe the sampling 
of featured and non-featured Wikipedia articles in more detail. 
Recall that our goal here is to create, for each featured article,
a comparison set of non-featured articles that have roughly
the same number of edits at roughly the same periods of time.

Our procedure is as follows.
For each article $a$, we let $e_b(a,y)$ be the number of times $a$ was edited before year $y$. Similarly, we let $e_d(a,y)$ and $e_a(a,y)$ be the number of times $a$ was edited during and after year $y$. For each article $a_f$ featured in year $y$, we find $k$ random non-featured articles, $L_{a,k}$, 
that have approximately the same number of edits as $a_f$ during the
years before and after $y$. That is, for an article $a_f$ featured
during year $y$, the non-feature article $a_n$ can be in $L_{a,k}$ if
$\frac{|e_b(a_f,y) - e_b(a_n,y)|}{e_b(a_f,y) } < .05$ and
$\frac{|e_a(a_f,y) - e_a(a_n,y)|}{e_a(a_f,y) } < .05$.

We aim to investigate the differences in the amount of coordination between featured and non-featured articles that are not a direct consequence of differences such as volumes of edits. Hence, we also require that $e_b(a_f,y) < e_b(a_n,y)$ for $a_n$ to be included in $L_{a,k}$. The results turn out to be the same with and without this additional restriction. We choose the article $a_n$ from the set of all non-featured articles without replacement. That is, the sets $L_{a,k}$ are pairwise disjoint. 

We define $F$ to be the set of articles that have been featured on Wikipedia for which $L_{a, k}$ in non-empty,
and we define 
$NF =  \cup_{a \in F}\{L_{a,30}\}$, the non-featured articles with approximately
the same number of edits as a featured article during the years before and after the year the article was featured. Throughout the paper we compare the sets $F$ and $NF$. 
We repeat the analysis with different choices of $k$ and find that the
results are consistent for all moderate values of $k$. We present the results for $k=30$

\bibliographystyle{aaai}
\bibliography{n}

\end{document}